\documentclass[useAMS,usenatbib]{mn2e}
\usepackage{graphicx}
\usepackage{txfonts}
\usepackage{natbib}

\title[30 day period in SAX~J1818.6$-$1703] {Discovery of a 30 day period in the supergiant fast X-ray transient SAX~J1818.6$-$1703}

\author[A.J.Bird et al.]{A. J. Bird$^{1}$, A. Bazzano$^{2}$, A.B. Hill$^{1}$, V.A. McBride$^{1}$, V. Sguera$^{3}$, S.E. Shaw$^{1}$, \newauthor H. J. Watkins$^{1}$ \\
$^{1}$ School of Physics and Astronomy, University of Southampton, SO17 1BJ, UK \\
$^{2}$ INAF-IASF, Via del Fosso del Cavaliere 100, 00133 Roma, Italy \\
$^{3}$ INAF-IASF, Via Gobetti 101, Bologna, Italy 
}

\begin{document}

\date{Accepted --- . Received --- ; in original form ---}

\pagerange{\pageref{firstpage}--\pageref{lastpage}} \pubyear{2008}

\maketitle

\label{firstpage}


\begin{abstract}

SAX J1818.6$-$1703 has been characterised as a Supergiant Fast X-ray Transient system on the basis of several INTEGRAL/IBIS detections since the original BeppoSAX Wide Field Camera detection. Using IBIS/ISGRI, Swift/BAT and archival observations, we show that in fact SAX J1818.6$-$1703 exhibits emission on a period of 30 $\pm$ 0.1 days, with a high degree of recurrence. SAX J1818.6$-$1703 is therefore the second SFXT shown to exhibit periodic outbursts, but with a considerably shorter period than the other known system, IGR J11215$-$5952. 

\end{abstract}



\section{Introduction}

In just over 5 years, the IBIS instrument \citep{2003A&A...411L.131U} on board the INTEGRAL gamma-ray satellite \citep{2003A&A...411L...1W} has revolutionised our classical view of High Mass X-ray binaries (HMXB). Specifically, IBIS has discovered many new HMXBs hosting massive OB supergiant stars (the so-called Supergiant High Mass X-ray binaries, SGXB); in just a few years their population has been almost tripled thanks to the IBIS identification of two previously unrecognized subclasses of SGXBs: the highly obscured SGXBs \citep{2005AIPC..797..402K} and the Supergiant Fast X-ray Transients (SFXTs; \citealt{2005AA...444..221S, 2006ApJ...646..452S}, \citealt{2006ESASP.604..165N}). 

Around 9 SFXTs are known to date. They spend most of the time in a low level of X-ray activity characterized by X-ray luminosity values in the range 10$^{32}$--10$^{34}$ erg s$^{-1}$, well below the persistent bright state of other classical SGXBs ($\sim$10$^{36}$ erg s$^{-1}$). Only occasionally do they display bright and fast X-ray outbursts with a duration no longer than a few days (typically only a few hours) and a dynamical range in flux of 10$^{3}$--10$^{4}$. A number of hypotheses have been proposed to explain the peculiar flaring behaviour of SFXTs, both as a class of objects on their own, and in the context of persistently accreting supergiant X-ray binaries.  Some of the models proposed invoke structure in the wind of the supergiant companion.  This structure could be either in the form of clumping of spherically symmetric outflow from the supergiant donor \citep{2005A&A...441L...1I,2007A&A...465L..35L,2007A&A...476..335W}, or in the form of an equatorially density enhanced wind from the supergiant, inclined at some angle to the orbit of the neutron star \citep{2007A&A...476.1307S}.  Futhermore, variations in the orbital eccentricity \citep{2008AIPC.1010..252N} can be used to explain the difference between supergiant X-ray binary systems showing transient as opposed to persistent X-ray emission.  A different model proposed by \cite{2008ApJ...683.1031B}  makes use of transitions between accretion gating mechanisms, such as centrifugal and magnetic barriers, brought about by variations in the stellar wind, to explain the large dynamic range in flux observed in these systems.   

Fast outbursts from SFXTs are relatively rare occurances (i.e. a few are detected per year), and long periods of inactivity are interspersed here and there with short flares spaced by irregular intervals of time. Because of this, the identification of periodicity is a very challenging task which can be fullfilled only through long-term and continous monitoring. To date, while recurrent (but non-periodic) outbursts have been observed in many SFXTs, only one, IGR J11215$-$5952 has been shown to exhibit periodic flaring activity separated by regular intervals of $\sim$ 165 days \citep{2006A&A...450L...9S, 2007ATel.1151....1R}. The identification of periodicity in the outburst behaviour of SFXTs,  most likely represents the orbital period of the binary system, and is a key diagnostic for studying the geometry of the system (i.e. orbital radius, eccentricity) and hence for understanding the physical reasons behind their very unusual X-ray behaviour.  In this letter we report a comprehensive temporal study of the SFXT SAX J1818.6$-$1703, leading to the discovery of a $\sim$ 30 day periodicity, making this the second confirmed periodic SFXT.

\section{Previous observations} \label{sec:previous}

SAX J1818.6$-$1703 was reported for the first time by BeppoSAX/WFC on March 11, 1998 during one of the Galactic Bulge observation campaigns \citep{1998IAUC.6840....2I}. The peak flux was $\sim$400 mCrab (9-25 keV) and the outburst lasted for 0.1 days and was characterised by a fast decay. 


On September  9, 2003, the IBIS telescope on INTEGRAL detected 2 short and intense outbursts with fluxes reaching 400 mCrab in the 18-45 keV energy band. This source activity was reported by \cite{2005AstL...31..672G} as showing a complex time profile lasting for about 1 day, while the main event duration was 2.7 hours. Spectra at different evolutionary phases extending up to $\sim$100 keV were fitted with thermal bremsstrahlung models. The nature of the source was still unclear but RXTE and INTEGRAL data exhibiting such outbursts were considered indicative of a possible SFXT \citep{2005AA...444..221S}, a picture later strengthened by \cite{2006ATel..831....1N} on the basis of an optical/NIR counterpart with an I-band spectrum typical of an OB supergiant. This result was confirmed soon afterwards by \cite{2006ATel..915....1I}, with a Chandra observation revealing only one bright and variable source within the error region of SAX J1818.6$-$1703, and 0.21'' and 0.47'' from the 2MASS and USNO-B1.0 positions of the proposed counterpart. The authors reported a strongly absorbed ($N_H \sim 6.0 \times 10^{22} {\rm cm}^{-2}$) power law spectrum with a photon index of 1.9 and a 0.5-10 keV flux of $7.5 \times 10^{-12} ~ {\rm erg~cm}^{-2}{\rm~s}^{-1}$ and also suggested the source may not be a true transient but a persistent HMXB with luminous flares such as 4U 1907+097 or Vela X-1. \cite{2008A&A...482..113M} confirmed the OB supergiant nature of the counterpart and inferred a distance of 2.5kpc.

The source was not detected by  XMM during a 13 ks observation on October 8,  2006. Following the new report of activity by \cite{2008ATel.1482....1G}, a Swift ToO was activated on April 18, 2008 and the flux value recorded was consistent with the Chandra  observation, while the XMM upper limit is a factor of 80 below these detections. With these new findings, \cite{2008ATel.1493....1B} concluded that SAX J1818.6$-$1703 shows an X-ray flux dynamic range similar to that observed for other SFXTs. 
 
\section{Dataset and analysis}

IBIS/ISGRI data is organised and analysed in short pointings (science windows, scw) of typically 2000s duration. We analysed IBIS/ISGRI data for all available science windows covering the date range from March 12, 2003 to March 12, 2008 (MJD range 52710-54537) in the 18-60 keV energy band using the standard OSA 7 analysis tools to provide intensity and significance images for each science window. We excluded data flagged as taken during bad time intervals, and additionally removed science windows with high image rms. Thereafter, we extracted fluxes from each science window at the position of SAX J1818.6$-$1703 to create a light curve at science window resolution. For periodicity analysis, science windows where SAX J1818.6$-$1703 was more than 12 degrees off-axis were excluded to avoid any effects due to incomplete off-axis response correction \citep{2007A&A...467..249S}. After filters were applied, the IBIS dataset consisted of 3756 scw for a total effective exposure of 6.1 Ms.

The Swift/BAT 15-50 keV light curve for SAX J1818.6$-$1703 was obtained from the BAT Transient Monitor programme \citep{2006HEAD....9.1347K}, and covered the date range from May 10, 2005 to August 23, 2008 (MJD range 53500-54705). Data were filtered to include only pointings with data quality flag 0 (good quality). After filtering, 10975 individual pointings with a total effective exposure of 5.6 Ms were retained.

\section{Periodicity analysis}

\begin{figure}
\includegraphics[width=\linewidth]{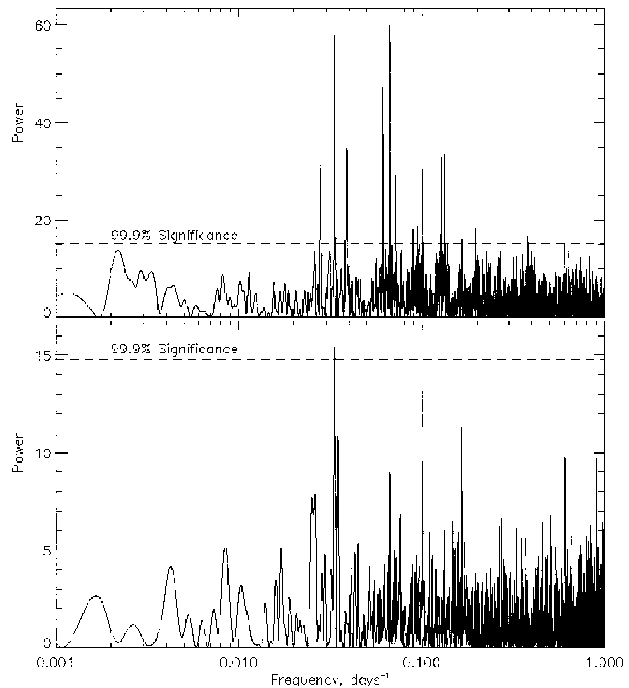}
\caption{Lomb-Scargle periodogram for the IBIS/ISGRI 18-60 keV (upper) and Swift/BAT 15-50 keV (lower) light curves. The 99.9\% significance power is 15.1 for IBIS and 14.8 for Swift/BAT, derived using the standard formulation \citep{1986ApJ...302..757H}\label{fig:LS}}
\end{figure}

A standard test for periodicity was carried out using the Lomb-Scargle method \citep{1976Ap&SS..39..447L,1982ApJ...263..835S,1989ApJ...343..874S} on the IBIS combined light curves in the 18-60 keV band.  An indication of periodic signal at around 30 days was found, but with a significant degree of aliasing and a strong first harmonic (Figure~\ref{fig:LS}, upper panel).
 
Some aliasing is not unusual in IBIS light curves, because of the window function imposed by the approximately 6-monthly observation pattern for sources near the Galactic Bulge. Furthermore, significant leakage of power into the harmonic is usually considered an indicator of non-sinusoidal modulation, which the Lomb-Scargle algorithm must compensate for by the addition of higher frequency components. To provide an independent verification of the suspected periodicity, the same analysis was performed on the Swift/BAT 15-50 keV light curve (Figure ~\ref{fig:LS}, lower panel). In this case, the aliasing is much reduced due to the more even temporal coverage provided by the large BAT field of view, but the power of the periodicity is substantially reduced due to the lower sensitivity per pointing compared to IBIS/ISGRI. Nevertheless, the periodicity at $\sim$30 days is still detected at above 99.9\% significance in both light curves.

\begin{figure}
\includegraphics[width=\linewidth]{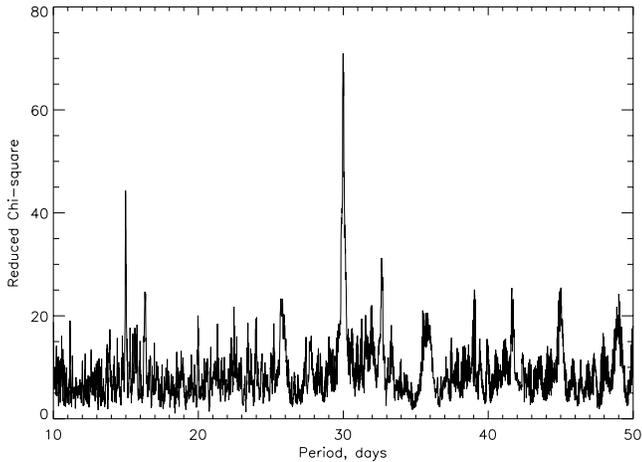}
\caption{Epoch folding analysis of the IBIS/ISGRI 18-60 keV lightcurve. \label{fig:EFsearch}}
\end{figure}

In order to investigate further, the detected periodicity was verified by the use of an epoch-folding method on the IBIS/ISGRI light curve, this method being less dependent on the periodic flux profile. This search again resulted in a clear maximum at around 30 days (Figure~\ref{fig:EFsearch}). Fitting the peak yields an estimate of both the period and its associated error, specifically $30.0 \pm 0.1$ (1$\sigma$) days.

\begin{figure}
\includegraphics[width=\linewidth]{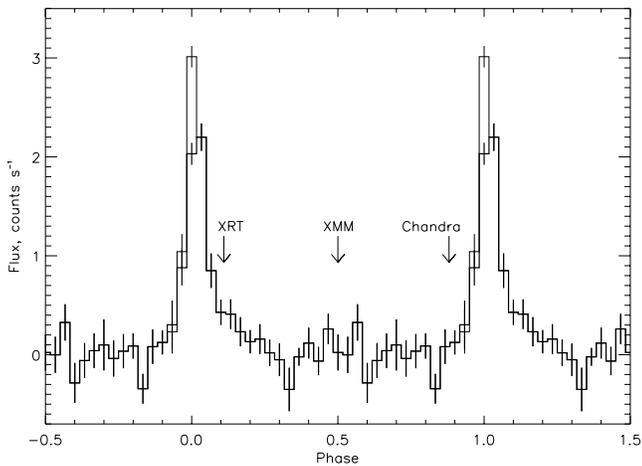}
\caption{IBIS light curve folded on the 30 day period. The narrow line is for the whole dataset, while the bold line is the result when the largest outburst is removed.The arrows refer to the X-ray measurements discussed in Section~\ref{section:discussion}.} \label{fig:folded}
\end{figure}

The IBIS light curve folded on the 30 day period (Figure~\ref{fig:folded}) shows a very significant modulation, and also shows that the source has a duty cycle of around 4-6 days during each 30 day period. Simulations performed on data with the same duty cycle verified that the strong harmonics observed in the Lomb-Scargle analysis are due to this profile.

In order to ensure that the large outburst at around MJD 52891 was not dominating the overall folded light curve, the analysis was repeated removing the largest outburst (specifically data for MJDs 52889-52893). Figure~\ref{fig:folded} also shows the folded IBIS light curve with the outburst removed, and there is no substantial difference in shape with or without the large outburst. Indeed, the IBIS light curve analysed represents some 70 cycles of the $\sim$30d period, and should be relatively robust against any such dominance by a single outburst. Furthermore, the analysis in section \ref{section:recurrence} shows that SAX J1818.6$-$1703 has a high degree of recurrence, indicating that this folded light curve should be a reasonable representation of the average outburst.

\section{Search for new outbursts}

Knowledge of the period and phasing of the known outbursts is of great benefit when examining the light curves for events of lower significance. A search was carried out for other outbursts with similar characteristics to the known ones in both the IBIS and BAT light curves. A significance was determined for a 4-day period around each of the expected outburst times where sufficient data were available. Table~\ref{table:newbursts} reports detections above 7 sigma in either instrument, and the corresponding significance in the other instrument lightcurve.

\begin{table*}
\caption{New outbursts from SAX J1818.6$-$1703 detected using the period determination \label{table:newbursts}}
\begin{tabular}{|l|c|c|c|c|c|c|} \hline
 & MJD & IBIS significance & IBIS pointings & BAT significance & BAT pointings \\ 
  &         &   (18-60 keV) & & (15-50 keV) & \\ \hline
A & 53279.8 - 53283.8 & 7.5 & 97 & no data & no data \\
B & 53669.4 - 53673.4 & 14.8 & 68 & 0.5 & 27 \\
C & 53849.3 - 53853.3 & 14.5 & 37 & 6.5 & 35 \\
D & 54388.8 - 54392.8 &  3.9 &  5   & 9.9 & 42 \\
\hline
\end{tabular}

\end{table*}

\begin{figure}
\includegraphics[width=\linewidth]{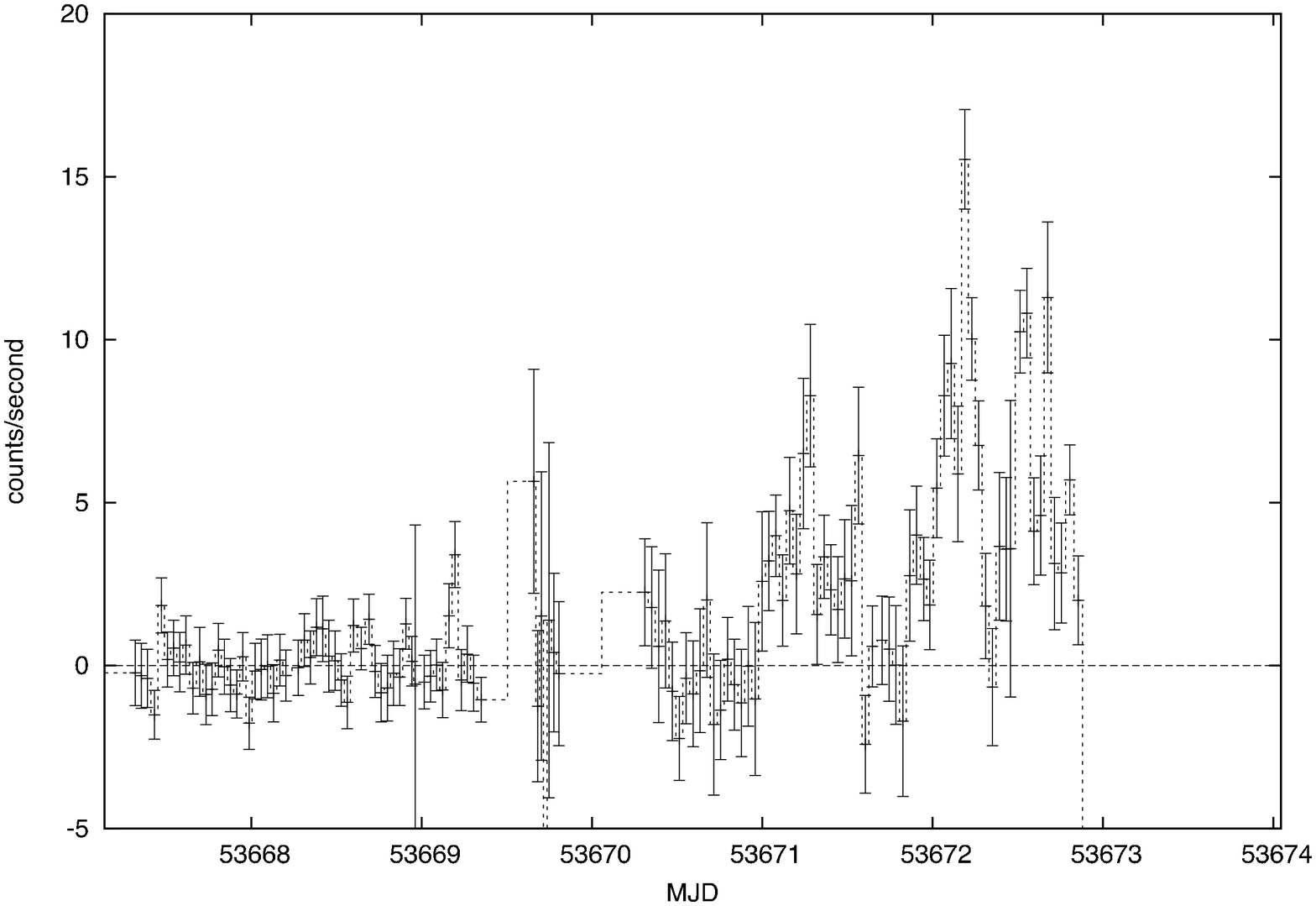}
\includegraphics[width=\linewidth]{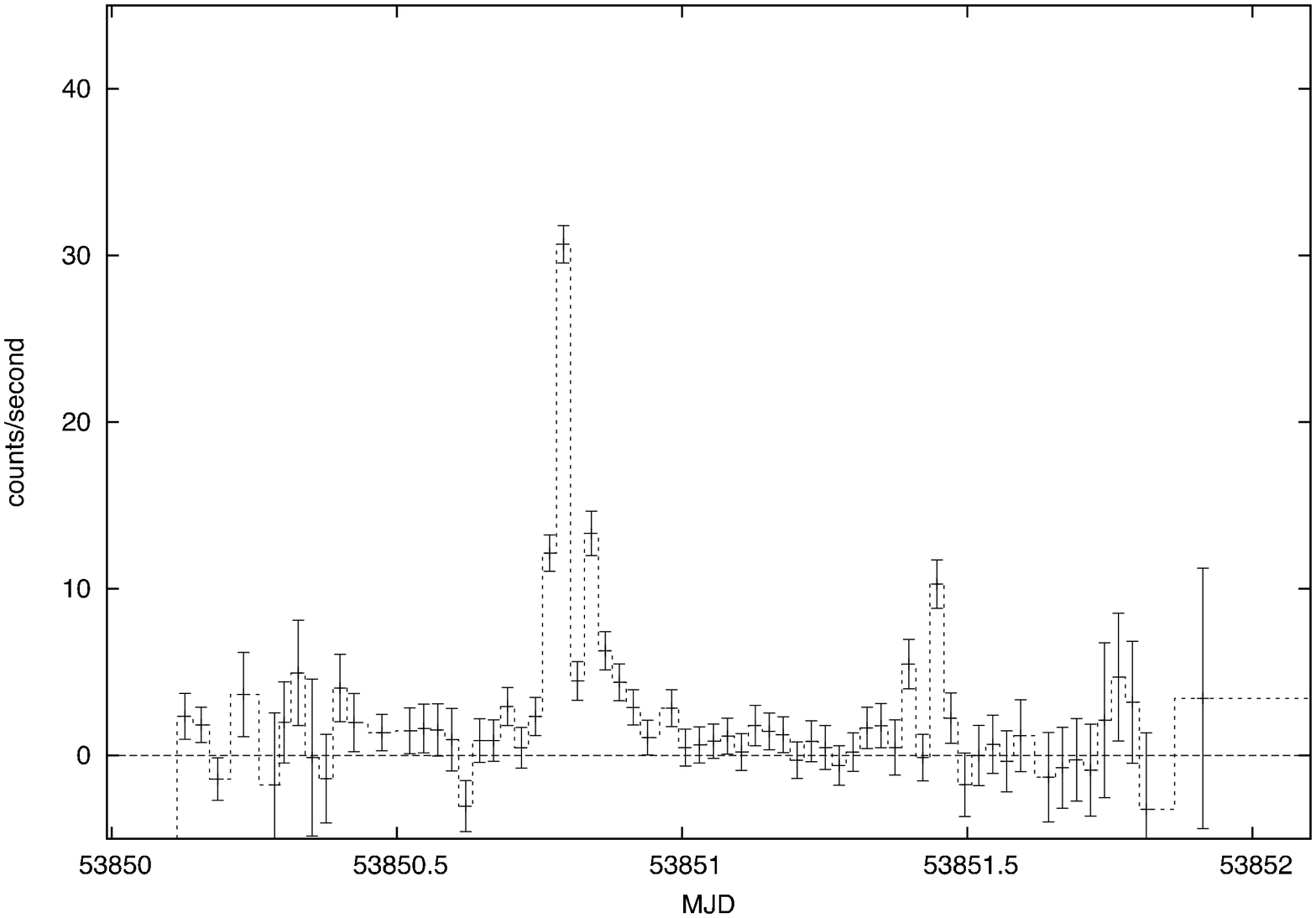}
\caption{IBIS 18-60 keV light curves showing the previously unreported flares at MJDs 53671 and 53851 \label{fig:newburst} (flares B and C in Table~\ref{table:newbursts}).}
\end{figure}

The IBIS 18-60 keV light curves of the newly discovered outbursts around MJDs 53671 and 53851 (labelled B and C in Table~\ref{table:newbursts}) are shown in Figure~\ref{fig:newburst}.

The period of flaring activity around MJD 53671 is similar to that observed in other SFXTs, a period of activity of around 2 days made up of several faster flares lasting a few hours. By contrast, the outburst detected around MJD 53851 shows a much cleaner structure, with just two detectable flares. Note that, apart from the different intensities of the outbursts (around 50 mCrab and 120 mCrab respectively) the data quality for the earlier flare is also degraded because the source is further from the telescope axis.

\section{Outburst Recurrence} \label{section:recurrence}

While only detections above 7 sigma are reported in Table~\ref{table:newbursts}, a number of other detections in the significance range 3-7 sigma were indicated during the search for new outbursts. This suggests that the recurrence rate for outbursts is quite high for SAX J1818.6-1703, and to investigate this further, the IBIS light curve was analysed to determine the 4-day significances both around, and well away from, the predicted outburst times.

\begin{figure}
\includegraphics[width=\linewidth]{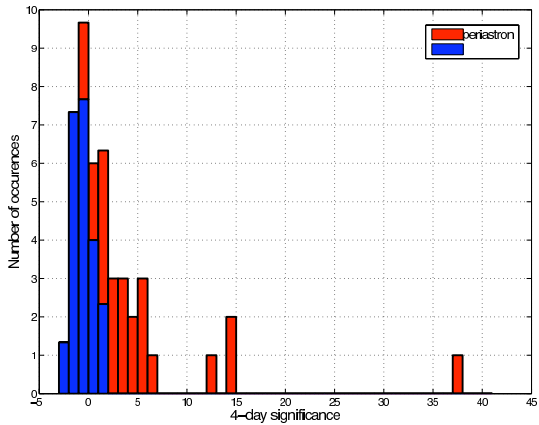}
\caption{Significance distributions for all available 4-day periods around periastron (red) and away from periastron (blue, scaled) \label{fig:recur}}
\end{figure}

For the IBIS light curve, data were available for 23 possible outburst times where more than two scw fell within the 4 day window, while 68 independent tests were performed for 4-day windows whose starts were offset by between 9 and 17 days after a predicted outburst. Of the 23 tests performed around the predicted outbursts times, 15 produced a significance above 3 sigma, while none of the 68 tests away from those times did. The distribution of 4-day IBIS significances is shown in Figure~\ref{fig:recur} and it is evident that a large fraction of the predicted outbursts produce emission above the quiescence level. Furthermore, at least two of the apparent non-detections may be due to limited exposure during potential outbursts, and so a conservative lower estimate can be made that detectable emission is seen on more than 50\% of the times when an outburst is predicted by the 30-day period.

\section{Discussion} \label{section:discussion}

This 30 day periodicity, evidenced by a regular increase in emission and occasional strong outbursts, is strongly indicative of a binary orbital period. The various observations of SAX J1818.6$-$1703 performed to date, placed in the context of this new orbital period, are summarised in Table~\ref{table:timeline}. This table includes both the INTEGRAL/IBIS and Swift/BAT detections, together with the X-ray observations performed by various telescopes. 

\begin{table*}
\caption{Timeline of X-ray and hard X-ray observations of SAX J1818.6$-$1703 \label{table:timeline}}
\begin{tabular}{|c|c|c|c|c|c|c|} \hline
Instrument & MJD &  Phase & IBIS peak flux & Notes & References \\ 
                    &          &               &   (18-60 keV, mCrab)  &             & \\
\hline
SAX/WFC & 50883.86  & 0.07 & - & Discovery & \cite{1998IAUC.6840....2I}\\
IBIS            & 52891.6    & 0.01 & 400 & Large outburst &  \cite{2005AstL...31..672G} \\
IBIS            & 52922.96  & 0.06 &   50     & Flaring activity & \cite{2005AA...444..221S} \\
IBIS            & 53281           & 1.00        &   $<$50     & New & \\
IBIS            & 53671          &  0.00       &   50     & New & \\
IBIS + BAT    & 53851    & 1.00 &   120     & New & \\
Chandra   & 53997.5   & 0.89 &      & Rising flux & \cite{2006ATel..915....1I} \\
XMM          & 54016     & 0.51 &       & Upper limit & \cite{2008ATel.1493....1B} \\
IBIS + BAT & 54390   & 0.98  &  $<$50     & New & \\
BAT            & 54540.659 & 1.00    &    & BAT trigger & \cite{2008GCN..7419....1B} \\
IBIS            & 54568      & 0.91 &       & Rising flux & \cite{2008ATel.1482....1G}\\
Swift/XRT            & 54574    & 0.11 &     & Detection  & \cite{2008ATel.1493....1B} \\
\hline
\end{tabular}

\end{table*}

The results of the X-ray observations carried out to date are consistent with this orbital period, and for reference the phase (arbitrarily referenced to the BAT trigger at MJD 54540.659) of the various X-ray observations are indicated in Figure~\ref{fig:folded}. It can be seen that the XMM observation, which reported a firm upper limit on the flux, occurred at a phase of 0.51, i.e. apastron. Thus this measurement can be interpreted as setting an upper limit on the quiescent flux from the system. By contrast, the Chandra observation was performed at phase 0.89, and the reported monotonic rising flux appears to show the onset of emission, but whether this is a slow increase as the compact object approached periastron, or the onset of an outburst, we cannot say. The XRT observation was performed at a phase of 0.11 and indicates some low-level activity, compatible with the Chandra flux, and an order of magnitude above the XMM upper limit.

We performed a similar analysis using the RXTE/ASM light curve. Unfortunately, while we can conclude that RXTE/ASM is detecting X-ray emission from the source, it does not provide any further diagnostic information due to its limited instantaneous sensitivity to a source that is only rarely above 20mCrab, and time coverage that does not really manage to capture the detail of the flaring time structure.

A systematic search of the IBIS data for both the large outburst and newly discovered outbursts has been performed to look for pulsations in the range 0.2 - 5000s. No statistically significant pulse period has been detected by this analysis and indeed no pulsations from SAX J1818.6$-$1703 have been reported in the literature.

We conclude that although SAX J1818.6$-$1703 was originally detected as a transient, and later categorised as a SFXT, knowledge of the orbital solution enables us to see a rather different picture. SAX J1818.6$-$1703 is a supergiant system with a compact object in a 30 day orbit, and shows regular emission on at least 50\% of the periastron passages. Moreover, some periastron passages give rise to larger outbursts, during which fast flaring structure is evident in the light curve - it is this structure which has led to the identification as a SFXT. SAX J1818.6$-$1703 shows an extreme dynamic range of emission, from $<1.1\times 10^{-13}~ {\rm erg.cm}^{-2}{\rm.s}^{-1}$ (0.5-10 keV) at apastron \citep{2008ATel.1493....1B}, to $3.7 \times 10^{-9}~ {\rm erg.cm}^{-2}{\rm.s}^{-1}$ by extrapolation of the best-fit spectrum during the largest recorded IBIS outburst \citep{2005AstL...31..672G} to 0.5-10keV, which again supports the SFXT categorisation. The simplest interpretation of this behaviour is of a compact object in an eccentric orbit, periodically accreting from the wind of a donor supergiant.

As the second SFXT to be shown to have a periodic outbursting behaviour, SAX J1818.6$-$1703 adds to our overall picture of the supergiant binary systems, placed between the persistent systems with circular orbits, and the long-period system IGR J11215$-$5952 which shows outbursts on a 165 day period. This indicates that orbital characteristics may play an important role in the accretion mechanisms of these systems, much as they do in Be transient systems. The shorter period of SAX J1818.6$-$1703 has enabled us to determine that the system shows a high degree of recurrence. The system appears similar to IGR J18483$-$0311, which has an orbital period of 18.52 days,  spin period of 21.0526 s \citep{2007A&A...467..249S}, and has recently been confirmed as a supergiant system \citep{2008arXiv0809.4415R}. Other SFXT systems may eventually reveal similar behaviour, but with more eccentric orbits and/or lower recurrence rates which have so far prevented the determination of their orbital periods. 

We encourage detailed follow-up observations based on this orbital period, such as have proved so effective for IGR J11215$-$5952. A detailed picture of the temporal structures seen during both faint and bright periastron passages should prove an invaluable probe into the environment from which the compact object is accreting.


\section*{acknowledgments}

\small

Based on observations with INTEGRAL, an ESA project with instruments and science data centre funded by ESA member states (especially the PI countries: Denmark, France, Germany, Italy, Switzerland, Spain), Czech Republic and Poland, and with the participation of Russia and the USA. HJW thanks the School of Physics and Astronomy, University of Southampton for financial support. We acknowledge the following funding: in Italy, ASI-INAF contract I/023/05/0 and I/088/06/0; in the UK, STFC grant GR/2002/00446; A.B. and V.S. acknowledge the support of Nature (455, 835-836) and thank the Editors for increasing the international awareness of the current critical situation in Italian Research.

\normalsize



\begin{thebibliography}{99}

\bibitem[\protect\citeauthoryear{Barthelmy et al.}{2008}]{2008GCN..7419....1B} 
Barthelmy S.~D., Krimm H.~A., Markwardt C.~B., Palmer D.~M., Ukwatta T.~N., 2008, GCN, 7419, 1

\bibitem[\protect\citeauthoryear{Bozzo et al.}{2008a}]{2008ATel.1493....1B}
Bozzo E., Campana S., Stella L., Falanga M., Israel G., Rampy R., Smith D., Negueruela I., 2008a, ATel, 1493, 1

\bibitem[\protect\citeauthoryear{Bozzo, Falanga, \& Stella}{2008b}]{2008ApJ...683.1031B} 
Bozzo E., Falanga M., Stella L., 2008b, ApJ, 683, 1031

\bibitem[\protect\citeauthoryear{Grebenev \& Sunyaev}{2005}]{2005AstL...31..672G} 
Grebenev S.~A., Sunyaev R.~A., 2005, AstL, 31, 672

\bibitem[\protect\citeauthoryear{Grebenev \& Sunyaev}{2008}]{2008ATel.1482....1G} 
Grebenev S.~A., Sunyaev R.~A., 2008, ATel, 1482, 1
 
 \bibitem[\protect\citeauthoryear{Horne \& Baliunas}{1986}]
{1986ApJ...302..757H} Horne J.~H., Baliunas S.~L., 1986, ApJ, 302, 757 

\bibitem[\protect\citeauthoryear{in 't Zand et al.}{1998}]{1998IAUC.6840....2I} 
in 't Zand J., Heise J., Smith M., Muller J.~M., Ubertini P., Bazzano A., 1998, IAUC, 6840, 2

\bibitem[\protect\citeauthoryear{in 't Zand}{2005}]{2005A&A...441L...1I} 
in 't Zand J.~J.~M., 2005, A\&A, 441, L1 
 
\bibitem[\protect\citeauthoryear{in 't Zand et al.}{2006}]{2006ATel..915....1I} 
in 't Zand J., Jonker P., Mendez M., Markwardt C., 2006, ATel, 915, 1
 
\bibitem[\protect\citeauthoryear{Krimm et al.}{2006}]{2006HEAD....9.1347K}
Krimm H.~A., Barthelmy S.~D., Markwardt C.~B., Sanwal D., Tueller J., Gehrels N., for the Swift/BAT Team, 2006, HEAD, 38, 374
 
\bibitem[\protect\citeauthoryear{Kuulkers}{2005}]{2005AIPC..797..402K}
Kuulkers E., 2005, AIPC, 797, 402
 
\bibitem[\protect\citeauthoryear{Leyder et al.}{2007}]{2007A&A...465L..35L} 
Leyder J.-C., Walter R., Lazos M., Masetti N., Produit N., 2007, A\&A, 465, L35
 
\bibitem[\protect\citeauthoryear{Lomb}{1976}]{1976Ap&SS..39..447L} 
Lomb N.~R., 1976, Ap\&SS, 39, 447
 
\bibitem[\protect\citeauthoryear{Masetti et al.}{2008}]{2008A&A...482..113M} 
Masetti N., et al., 2008, A\&A, 482, 113

\bibitem[\protect\citeauthoryear{Negueruela et al.}{2006}]{2006ESASP.604..165N} 
Negueruela I., Smith D.~M., Reig P., Chaty S., Torrej{\'o}n J.~M., 2006, ESASP, 604, 165
 
 
\bibitem[\protect\citeauthoryear{Negueruela \& Smith}{2006}]{2006ATel..831....1N} 
Negueruela I., Smith D.~M., 2006, ATel, 831, 1
 
 
\bibitem[\protect\citeauthoryear{Negueruela et al.}{2008}]{2008AIPC.1010..252N} 
Negueruela I., Torrej{\'o}n J.~M., Reig P., Rib{\'o} M., Smith D.~M., 2008, AIPC, 1010, 252

\bibitem[\protect\citeauthoryear{Rahoui \& Chaty}{2008}]{2008arXiv0809.4415R} 
Rahoui F., Chaty S., 2008, arXiv, arXiv:0809.4415 
 
\bibitem[\protect\citeauthoryear{Romano et al.}{2007}]{2007ATel.1151....1R}
Romano P., Mangano V., Mereghetti S., Paizis A., Sidoli L., Vercellone S., 2007, ATel, 1151, 1
 
 
\bibitem[\protect\citeauthoryear{Scargle}{1982}]{1982ApJ...263..835S}
Scargle J.~D., 1982, ApJ, 263, 835
 
 
\bibitem[\protect\citeauthoryear{Scargle}{1989}]{1989ApJ...343..874S}
Scargle J.~D., 1989, ApJ, 343, 874
 
 
\bibitem[\protect\citeauthoryear{Sguera et al.}{2005}]{2005AA...444..221S} 
Sguera V., et al., 2005, A\&A, 444, 221
 
 
\bibitem[\protect\citeauthoryear{Sguera et al.}{2006}]{2006ApJ...646..452S}
Sguera V., et al., 2006, ApJ, 646, 452
 
 
\bibitem[\protect\citeauthoryear{Sguera et al.}{2007}]{2007A&A...467..249S} 
Sguera V., et al., 2007, A\&A, 467, 249
 
 
\bibitem[\protect\citeauthoryear{Sidoli, Paizis, \& Mereghetti}{2006}]{2006A&A...450L...9S} 
Sidoli L., Paizis A., Mereghetti S., 2006, A\&A, 450, L9
 
 
\bibitem[\protect\citeauthoryear{Sidoli et al.}{2007}]{2007A&A...476.1307S} 
Sidoli L., Romano P., Mereghetti S., Paizis A., Vercellone S., Mangano V., G{\"o}tz D., 2007, A\&A, 476, 1307
 
 
\bibitem[\protect\citeauthoryear{Ubertini et al.}{2003}]{2003A&A...411L.131U} 
Ubertini P., et al., 2003, A\&A, 411, L131
 
 
\bibitem[\protect\citeauthoryear{Walter \& Zurita Heras}{2007}]{2007A&A...476..335W} 
Walter R., Zurita Heras J., 2007, A\&A, 476, 335
 
 
\bibitem[\protect\citeauthoryear{Winkler et al.}{2003}]{2003A&A...411L...1W} 
Winkler C., et al., 2003, A\&A, 411, L1
 

\end{thebibliography}
\end{document}